\documentclass[aps, twocolumn, letterpaper, superscriptaddress]{revtex4}

\usepackage{amsmath}
\usepackage{amssymb}
\usepackage{mathrsfs}
\usepackage[dvipdfmx]{graphicx}
\usepackage{color}
\usepackage{xspace}
\usepackage{bbold}
\newcommand{\shout}[1]{\textcolor{black}{#1}}

\newcommand{\eq}[1]{(\ref{#1})}
\newcommand{\Eq}[1]{Eq.~(\ref{#1})}
\newcommand{\Eqs}[1]{Eqs.~(\ref{#1})}
\newcommand{\Fig}[1]{Fig.~\ref{#1}}
\newcommand{\Figs}[1]{Figs.~\ref{#1}}
\newcommand{\Sec}[1]{Sec.~\ref{#1}}
\newcommand{\Ref}[1]{Ref.~\cite{#1}}
\newcommand{\Refs}[1]{Refs.~\cite{#1}}

\newcommand{\mc}[1]{\mathcal{#1}}
\newcommand{\mcc}[1]{\mathfrak{#1}}
\newcommand{\msf}[1]{\mathsf{#1}}

\newcommand{\pd}{\partial}
\newcommand{\del}{\nabla}
\renewcommand{\vec}[1]{\boldsymbol{\rm #1}}
\newcommand{\oper}[1]{\smash{\widehat{#1}}}
\newcommand{\dd}{\mathrm{d}}

\newcommand{\Deff}{\msf{D}}
\newcommand{\sigmap}{{\bar{\sigma}}}
\newcommand{\dk}{\pi}

\newcommand{\parade}{\textit{\mbox{PARADE}}\xspace}

\begin{document}

\title{Quasioptical modeling of wave beams with and without mode conversion:\\III. Numerical simulations of mode-converting beams}

\author{K. Yanagihara}
\affiliation{Nagoya University, 464-8601, Nagoya, Aichi, Japan}

\author{I. Y. Dodin}
\affiliation{Princeton Plasma Physics Laboratory, Princeton, New Jersey 08543, USA}

\author{S. Kubo}
\affiliation{Nagoya University, 464-8601, Nagoya, Aichi, Japan}
\affiliation{National Institute for Fusion Science, National Institutes of Natural Sciences, 509-5292, Toki, Gifu, Japan}

\date{\today}

\begin{abstract}
This work continues a series of papers where we propose an algorithm for quasioptical modeling of electromagnetic beams with and without mode conversion. The general theory was reported in the first paper of this series, where a parabolic partial differential equation was derived for the field envelope that may contain one or multiple modes with close group velocities. In the second paper, we presented a corresponding code \parade (PAraxial RAy DEscription) and its test applications to single-mode beams. Here, we report quasioptical simulations of mode-converting beams for the first time. We also demonstrate that \parade can model splitting of two-mode beams. The numerical results produced by \parade show good agreement with those of one-dimensional full-wave simulations and also with conventional ray tracing \shout{(to the extent that one-dimensional and ray-tracing simulations are applicable).}
\end{abstract}

\maketitle
\bibliographystyle{full}

\section{Introduction}
\label{sec:intro}

Geometrical-optics (GO) ray tracing has been widely used to calculate the propagation and absorption of electron-cyclotron waves (ECWs) in inhomogeneous magnetized fusion plasmas in many contexts, including electron-cyclotron heating, electron-cyclotron current drive, and electron-cyclotron emission diagnostics \cite{ref:tsujimura15, ref:marushchenko14}. Also, a lot of alternatives \cite{ref:mazzucato89, ref:nowak93, ref:peeters96, ref:farina07, ref:pereverzev98, ref:poli01b, ref:poli01, ref:poli18, ref:balakin08b, ref:balakin07a, ref:balakin07b}, which take into account diffraction to describe the beam width of ECWs near the focal region, have been proposed as extensions of the GO approach and have contributed to the improvement of the wave-power deposition-profile simulations. These approaches can treat single-mode waves in sufficiently dense plasmas. However, waves propagating in low-density plasmas often contain two electromagnetic modes, which have close refraction indexes and thus can interact efficiently. Specifically, in fusion plasmas, the mode conversion between the O and X waves that form ECWs is caused by the magnetic-field shear at the peripheral region. One of the ways to properly model this process is to perform one-dimensional full-wave (1DFW) analysis, such as that carried out in \Ref{ref:kubo15}. However, one-dimensional analysis becomes inaccurate already when the wave beams experience substantial bending. This can be remedied to some extent by applying ``extended geometrical optics'' (XGO) \cite{my:xo, phd:ruiz17, my:covar, my:qdirac, my:qdiel}, which is a reduced theory that describes refraction and mode conversion on the same footing. However, XGO, as it is formulated in \Refs{my:xo, phd:ruiz17, my:covar, my:qdirac, my:qdiel}, still does not include diffraction, so the problem remains open.

The present work continues a series of papers where we propose how to overcome this problem. \shout{Our results are quite general; in fact, they are not restricted even to plasma physics, let alone fusion applications.} The comprehensive theory generalizing XGO to include diffraction was reported in Paper~I of this series \cite{foot:paper1}. In Paper~II \cite{foot:paper2}, we presented a corresponding quasioptical code \parade (PAraxial RAy DEscription) in its reduced version that can simulate single-mode beams without mode conversion. Here, we present a more general version of \parade and report the first quasioptical simulations of mode-converting beams. We also demonstrate that \parade can model splitting of two-mode beams.

\shout{Although our work was primarily motivated by the need to improve ECW modeling in fusion plasmas, \parade can also be useful for studying related problems in optics and general relativity \cite{ref:bliokh15, foot:marius}. For this reason, we illustrate the \parade's capabilities below on basic-physics examples. (Fusion-relevant applications of \parade are discussed in \Ref{foot:itc27}). As seen from these examples, the numerical results produced by \parade are in good agreement with the corresponding results of ray-tracing and 1DFW simulations to the extent that those are applicable and such comparison is meaningful. More generally, \parade's simulations surpass ray tracing in that they resolve diffraction and mode conversion. \parade's simulations can also surpass 1DFW results in that they capture the true three-dimensional structure of wave beams, while the 1DFW approach is restricted to straight rays.}

Our paper is organized as follows. In \Sec{sec:xgo}, we introduce the key equations derived in Paper~I and also adjust them to numerical modeling. In \Sec{sec:sim}, we report simulation results for test problems. In \Sec{sec:conc}, we summarize our main conclusions.

\section{Theoretical model}
\label{sec:xgo}

\subsection{Basic equations}
\label{sec:basic}

Here, we outline how the general theory developed in Paper~I can be applied, with some adjustments, to describe mode-converting \shout{wave beams}. The general idea is similar to that presented in Paper~II for single-mode waves. We assume a general linear equation for the electric field~$\vec{E}$ of a wave,
\begin{gather}\label{eq:E}
\oper{\vec{D}} \vec{E} = 0,
\end{gather}
where $\oper{\vec{D}}$ is a linear dispersion operator. We also assume that this field can be represented in the eikonal form,
\begin{gather}\label{eq:psi}
\vec{E} = \vec{\psi} e^{i\theta},
\end{gather}
where $\vec{\psi}$ is a slow complex vector envelope and $\theta$ is a fast real ``reference phase'' to be prescribed. The wave is considered stationary, so it has a constant frequency $\omega$; then $\vec{\psi}$ and $\theta$ are functions of the spatial coordinate $\vec{x}$. Correspondingly, the envelope $\vec{\psi}$ satisfies
\begin{gather}\label{eq:mcD}
\oper{\vec{\mc{D}}} \vec{\psi} = 0,
\quad
\oper{\vec{\mc{D}}} \doteq e^{-i\theta(\vec{x})} \oper{\vec{D}} e^{i\theta(\vec{x})},
\end{gather}
where $\oper{\vec{\mc{D}}}$ serves as the ``envelope dispersion operator'' ($\doteq$ denotes definitions). We introduce $\vec{k} \doteq \del \theta (\vec{x})$ for the local wave vector, $\lambda \doteq 2\pi/k$ for the corresponding wavelength, $L_\parallel$ for the inhomogeneity scale of $\vec{\psi}$ along the beam, and $L_\perp$ for the minimum scale of $\vec{\psi}$ across the beam. The medium-inhomogeneity scale is assumed to be larger than or comparable with $L_\parallel$, and we adopt
\begin{gather}\label{eq:eps}
\epsilon_\parallel \doteq \lambda/L_\parallel,
\quad
\epsilon_\perp \doteq \lambda/L_\perp,
\quad
\epsilon_\parallel \sim \epsilon_\perp^2 \ll 1.
\end{gather}
Then, \Eq{eq:mcD} becomes
\begin{gather}\label{eq:psipsi}
\vec{\msf{D}}\vec{\psi} +  \oper{\vec{\mc{L}}}_\epsilon\vec{\psi}= 0,
\end{gather}
where $\vec{\msf{D}}$ serves as the ``effective dispersion tensor'' found from $\oper{\vec{D}}$ and the operator $\oper{\vec{\mc{L}}}_\epsilon = O(\epsilon_\perp)$ is specified in Paper~I (also see below). We suppose the following ordering:
\begin{gather}\label{eq:DHDA}
\vec{\Deff}_H = O(1), \quad \vec{\Deff}_A = O(\epsilon_\parallel),
\end{gather}
where the indices $H$ and $A$ denote the Hermitian and anti-Hermitian parts of $\oper{\vec{D}}$, respectively. Assuming that the spatial dispersion is weak, $\vec{\Deff}$ can be replaced with the homogeneous-plasma dispersion tensor,
\begin{gather}\label{eq:D}
\vec{D}(\vec{x}, \vec{p})
 = \frac{c^2}{16\pi \omega^2}\,[\vec{p}\vec{p} - (\vec{p} \cdot \vec{p})\mathbb{1}]
 + \frac{1}{16\pi}\,\vec{\varepsilon}(\vec{x}, \vec{p}),
\end{gather}
where $\mathbb{1}$ is a unit matrix and $\vec{\varepsilon}$ is the homogeneous-plasma dielectric tensor \cite{book:stix}; its dependence on $\omega$ is assumed but not emphasized, since $\omega$ is constant. (Here, $\vec{p}$ denotes any given wave vector, as opposed to $\vec{k}$, which is the specific wave vector determined by $\theta$; see above.)  Note that \Eq{eq:D} assumes the Euclidean metric. Although we shall also use curvilinear coordinates below, expressing $\vec{D}$ in those coordinates will not be necessary as explained in Sec.~VI\,D of Paper~I.

With $\vec{\msf{D}} \approx \vec{D}$ and $\vec{D}$ given by \Eq{eq:D}, \Eq{eq:DHDA} implies $\vec{\varepsilon}_A = O(\epsilon_\parallel)$. Hence, $\vec{D}_H$ is the dominant part of $\vec{D}$, and \Eq{eq:psipsi} yields
\begin{gather}\label{eq:Dpsi}
\vec{D}_H \vec{\psi} = O(\epsilon_\perp).
\end{gather}
Since the Hermitian matrix $\vec{D}_H = O(1)$ is the dominant part of $\vec{D}$ and has enough eigenvectors $\vec{\eta}_s$ to form a complete orthonormal basis, it is convenient to represent the envelope $\vec{\psi}$ in this basis,
\begin{gather}\label{eq:etaa}
\vec{\psi} = \vec{\eta}_s a^s,
\end{gather}
where $a^s$ are the complex amplitudes. [Summation over repeating indices is assumed. For all functions derived from $\vec{D}$, such as $\vec{\eta}_s$, the notation convention $f \equiv f(\vec{x}) \equiv f(\vec{x}, \vec{k}(\vec{x}))$ will also be assumed by default.] Then,
\begin{gather}\label{eq:Degn}
\vec{D}_H \vec{\psi} =  \vec{\eta}_s \Lambda_s a^s,
\quad
\vec{D}_H \vec{\eta}_s =  \Lambda_s \vec{\eta}_s\shout{,}
\end{gather}
where $\Lambda_s$ are the corresponding eigenvalues. Due to \Eq{eq:Dpsi} and the mutual orthogonality of all $\vec{\eta}_s$, this means that $\Lambda_s a^s$ is small for every given $s$ individually; hence, either $a^s$ is small or $\Lambda_s$ is small. In the latter case, $\vec{\eta}_s$ approximately satisfies the eigenmode equation, $\vec{D}_H\vec{\eta}_s = \Lambda_s \vec{\eta}_s \approx 0$, so it can be viewed as the local polarization vector of a GO mode. Then, the corresponding $a^s$ can be understood as the local scalar amplitude of an actual GO mode, and $a^s = O(1)$ is allowed.

\subsection{Polarization matrices}
\label{sec:pol}

In this paper, we consider the situation when there are \textit{two} modes, henceforth called O and X modes, whose eigenvalues $\Lambda_{\rm o} \approx \Lambda_{\rm x}$ are \textit{both} small within the region of interest. (This assumption is specified in \Sec{sec:qo}. Also note that the single-mode case is discussed in Paper~II.) In other words, the two modes are approximately in resonance with each other. Then,
\begin{gather}\label{eq:psia}
\vec{\psi} = \vec{\eta}_{\rm o} a^{\rm o} + \vec{\eta}_{\rm x} a^{\rm x} +
\bar{\vec{\eta}}\bar{a},
\end{gather}
where $\vec{\eta}_{\rm o}$ and $\vec{\eta}_{\rm x}$ are the O- and X-mode polarization vectors, $\bar{\vec{\eta}}$ is some third eigenvector of $\vec{D}$ that is orthogonal to both of them, and
\begin{gather}
a^{\rm o} = O(1), \quad a^{\rm x} = O(1), \quad \bar{a} = O(\epsilon_\perp).
\end{gather}
The small amplitude $\bar{a}$ can be easily calculated as a perturbation and is included in our theory \cite{foot:paper1} but does not need to be considered below explicitly. Instead, we introduce a two-dimensional amplitude vector
\begin{gather}
\vec{a} = \left(
\begin{array}{c}
a^{\rm o}\\
a^{\rm x}
\end{array}
\right)
\end{gather}
and the $3 \times 2$ ``polarization matrix'' $\vec{\Xi}$ that contains the vectors $\vec{\eta}_{\rm o}$ and $\vec{\eta}_{\rm x}$ as its columns,
\begin{gather}
\vec{\Xi} = \left(
\begin{array}{cc}
\vec{\eta}_{\rm o} & \vec{\eta}_{\rm x}
\end{array}
\right).
\end{gather}
Then, $\vec{\psi}$ can be expressed as follows:
\begin{gather}\label{eq:psiXi}
\vec{\psi} = \vec{\Xi} \vec{a} + O(\epsilon_\perp).
\end{gather}

We also introduce the \textit{dual} basis vectors $\vec{\eta}^{\rm o}$ and $\vec{\eta}^{\rm x}$ and an auxiliary polarization matrix
\begin{gather}
\vec{\Xi}^+ = \left(
\begin{array}{c}
\vec{\eta}^{{\rm o}*} \\ \vec{\eta}^{{\rm x}*}
\end{array}
\right).
\end{gather}
As seen easily, this matrix satisfies $\vec{\Xi}^+ \vec{\Xi} = \mathbb{1}$, and
\begin{gather}\label{eq:Lambda}
\vec{\Lambda} \doteq \vec{\Xi}^+ \vec{D}_H \vec{\Xi} = \left(
\begin{array}{cc}
\Lambda_{\rm o} & 0\\
0 & \Lambda_{\rm x}
\end{array}
\right).
\end{gather}
Also, one can express the amplitude vector as
\begin{gather}\label{eq:apsi}
\vec{a} = \vec{\Xi}^+ \vec{\psi}
\end{gather}
[here, there is no $O(\epsilon_\perp)$ correction, unlike in \Eq{eq:psiXi}], whose squared length $|\vec{a}|^2 \equiv |a^{\rm o}|^2 + |a^{\rm x}|^2$  satisfies
\begin{gather}
|\vec{a}|^2 \approx |\vec{\psi}|^2 \equiv |\psi^x|^2 + |\psi^y|^2 + |\psi^z|^2
\end{gather}
up to $O(\epsilon_\parallel)$. Since we consider the beam dynamics in coordinates that are close to Euclidean, it will be sufficient, within the accuracy of our model, to adopt \cite{foot:paper1}
\begin{gather}
\vec{\eta}^{\rm o} \approx \vec{\eta}_{\rm o}
\quad
\vec{\eta}^{\rm x} \approx \vec{\eta}_{\rm x}.
\end{gather}
Then, $\vec{\Xi}^+$ is simply the Hermitian conjugate of $\vec{\Xi}$.

\subsection{Reference ray and new coordinates}
\label{sec:rr}

We consider the wave evolution in curvilinear coordinates that are linked to a ``reference ray'' (RR), which is governed by
\begin{gather}\label{eq:rtpX}
\frac{\dd X^\alpha}{\dd\zeta} = \frac{V^{\alpha}_\star}{V_\star},
\quad
\frac{\dd K_\alpha}{\dd\zeta} = - \frac{1}{V_\star}\,\frac{\pd H_\star}{\pd X^\alpha}.
\end{gather}
Here, $\zeta$ is the path along the ray, $\vec{X}$ and $\vec{K}$ are the RR coordinate and wave vector,
\begin{gather}
V_\star^\alpha(\zeta) \doteq \frac{\pd H(\vec{X}, \vec{K})}{\pd K_\alpha}
\end{gather}
is the group velocity, $V_\star \doteq |\vec{V}_\star|$, and the index $\star$ denotes that the corresponding quantity is evaluated on $(\vec{X}, \vec{K})$. In particular, $H_\star \doteq H(\vec{X}, \vec{K})$, and $H$ is defined as follows:
\begin{gather}
H(\vec{x}, \vec{p}) \doteq \frac{1}{2}\,[\Lambda_{\rm o}(\vec{x}, \vec{p}) + \Lambda_{\rm x}(\vec{x}, \vec{p})].
\end{gather}
We require $H_\star$ to be exactly zero initially, which is ensured by choosing an appropriate $\vec{K}$; then, $H_\star$ remains zero at all $\zeta$, as seen from \Eqs{eq:rtpX}.

The RR-based coordinates are introduced as $\tilde{x} \equiv \{\zeta, \tilde{\varrho}^1, \tilde{\varrho}^2\}$, where $\tilde{\varrho}^\sigma$ are orthogonal coordinates transverse to the RR as specified in Paper~II. (Here and further, the indices $\sigma$ and $\sigmap$ span from 1 to 2; other Greek indices span from 1 to 3.) The basis vectors $\tilde{\vec{e}}_\mu$ of the new coordinates ($\dd \vec{x} = \tilde{\vec{e}}_\mu \dd\tilde{x}^\mu$) are defined such that
\begin{gather}\label{eq:eee}
\tilde{\vec{e}}_{\star\mu} \cdot \tilde{\vec{e}}_{\star\nu} = \delta_{\mu\nu},
\quad
\left[\pd \tilde{\vec{e}}_{\sigma}(\tilde{x})/\pd \tilde{\varrho}^\sigmap\right]_\star = 0.
\end{gather}
Then,
\begin{gather}
\vec{x} \approx \vec{X}(\zeta) + \left(
\begin{array}{cc}
\tilde{\vec{e}}_{\star 1} & \tilde{\vec{e}}_{\star 2}
\end{array}
\right)
\left(
\begin{array}{c}
\tilde{\varrho}^1\\
\tilde{\varrho}^2
\end{array}
\right).
\end{gather}
For the transformation matrices defined as
\begin{gather}
 \mc{X}^\alpha{}_\mu \doteq \frac{\pd x^\alpha}{\pd \tilde{x}^\mu} = (\tilde{\vec{e}}_\mu)^\alpha,
 \\
 \tilde{\mc{X}}^\mu{}_\alpha \doteq \frac{\pd \tilde{x}^\mu}{\pd x^\alpha} = (\vec{\mc{X}}^{-1})^\mu{}_\alpha,
\end{gather}
this leads to
\begin{gather}
\mc{X}^\alpha{}_\zeta \approx \mc{X}^\alpha_\star{}_\zeta + [(\tilde{\vec{e}}_{\star\sigma})^\alpha]'\tilde{\varrho}^\sigma,
\\
\mc{X}^\alpha{}_\sigma \approx \mc{X}^\alpha_\star{}_\sigma,
\\
\mc{X}^\alpha_\star{}_\mu = (\tilde{\vec{e}}_{\star\mu})^\alpha\shout{.}
\end{gather}
The specific choice of $\tilde{\vec{e}}_{\star\mu}$ is described in Paper~II.

We shall use tilde to denote the components of vectors and tensors measured in the RR-based coordinates~$\tilde{x}$. These components can be mapped to those in the laboratory coordinates using the standard formulas \cite{foot:paper2}. For example, for any vector or covector $\vec{A}$, one has
\begin{gather}
A^\alpha = \mc{X}^\alpha{}_\mu \tilde{A}^\mu,
\quad
A_\alpha = \tilde{A}_\mu \tilde{\mc{X}}^\mu{}_\alpha.
\end{gather}
We also introduce first-order partial derivatives
\begin{gather}
f_{|\alpha} \doteq \frac{\pd f(\vec{x}, \vec{k})}{\pd x^\alpha},
\quad
f^{|\alpha} \doteq \frac{\pd f(\vec{x}, \vec{k})}{\pd k_\alpha},
\end{gather}
and the second-order derivatives are denoted as follows:
\begin{gather}\notag
f_{|\alpha\beta} \doteq \frac{\pd^2 f(\vec{x}, \vec{k})}{\pd x^\alpha\pd x^\beta},
\kern 5pt
f^{|\alpha\beta} \doteq \frac{\pd^2 f(\vec{x}, \vec{k})}{\pd k_\alpha\pd k_\beta},
\kern 5pt
f^{|\alpha}_{|\beta} \doteq \frac{\pd^2 f(\vec{x}, \vec{k})}{\pd k_\alpha\pd x^\beta}.
\end{gather}
Accordingly, $\tilde{f}_{|\mu} \doteq \pd f(\vec{x}, \vec{k})/\pd \tilde{x}^\mu$, and so on.

\subsection{Quasioptical equation}
\label{sec:qo}

Let us split the matrix \eq{eq:Lambda} into its scalar part $H$ and its traceless part $\vec{M}$; this gives $\vec{\Lambda} = H \mathbb{1} + \vec{M}$. We assume $\vec{M}(\vec{x}, \vec{k}(\vec{x})) \lesssim O(\epsilon_\perp)$ and $\vec{M}_\star \lesssim O(\epsilon_\parallel)$; then,
\begin{gather}
\vec{M}(\vec{x}, \vec{k}(\vec{x}))
= \vec{M}_\star + \tilde{\vec{M}}_{\star|\sigma}\tilde{\varrho}^\sigma + \tilde{\vec{M}}{}_\star^{|\mu}\tilde{\dk}_\mu + o(\epsilon_\parallel),
\end{gather}
where $\tilde{\dk}_\mu \doteq \tilde{k}_\mu(\vec{x}) - \tilde{K}_\mu(\zeta)$ is given by $\tilde{\dk}_\mu = - \tilde{\varrho}^\sigma \tilde{H}_{\star|\sigma} \tilde{V}_{\star\mu}/V_\star^2$ \cite{foot:paper2}. We can also rewrite this as
\begin{gather}
\vec{M}(\vec{x}, \vec{k}(\vec{x}))
= \vec{M}_\star + \vec{M}_{\star|\alpha}\varrho^\alpha + \vec{M}{}_\star^{|\beta}\dk_\beta
 + o(\epsilon_\parallel),\\
\varrho^\alpha \doteq \mc{X}_\star^\alpha{}_{\sigma} \tilde{\varrho}^\sigma,
\quad
\dk_\beta = - \varrho^\alpha H_{\star|\alpha} V_{\star\beta}/V_\star^2.
\end{gather}
This leads to
\begin{gather}\notag
\vec{\Lambda}(\vec{x}, \vec{k}(\vec{x}))
\approx H \mathbb{1} + \vec{M}_\star + \tilde{\vec{\mcc{M}}}_{\star\sigma} \tilde{\varrho}^\sigma,\\
\tilde{\vec{\mcc{M}}}_{\star\sigma} \doteq (\vec{M}_{\star|\alpha} - \vec{M}{}_\star^{|\beta} H_{\star|\alpha} V_{\star\beta}/V_\star^2)
\mc{X}_\star^\alpha{}_{\sigma}.
\end{gather}
Let us also introduce the matrices $\vec{u}^\alpha \doteq \vec{M}^{|\alpha} \lesssim O(\epsilon_\perp)$,
\begin{gather}
\tilde{\vec{u}}^\sigma = \tilde{\mc{X}}^\sigma{}_\alpha \vec{u}^\alpha
\approx \tilde{\vec{u}}^\sigma_\star = \tilde{\mc{X}}^\sigma_\star{}_\alpha \vec{M}^{|\alpha}_\star,
\end{gather}
and a rescaled amplitude vector
\begin{gather}
\vec{\phi} \doteq \sqrt{V_\star} \vec{a}.
\end{gather}
By combining the results of Papers~I and II, one readily finds that $\vec{\phi}$ is governed by the following parabolic partial differential equation, which is the main ``quasioptical'' equation used in \parade:
\begin{align}
V_\star \pd_\zeta \vec{\phi} =
& - (\tilde{\vec{u}}^\sigma_\star + \tilde{\vec{\vartheta}}_\star^\sigma{}_{\sigmap} \tilde{\varrho}^{\sigmap}) \pd_\sigma \vec{\phi}
- \frac{\tilde{\vec{\vartheta}}_\star^\sigma{}_\sigma}{2} \vec{\phi} \notag
\\ & + \frac{i}{2}\,\tilde{\vec{\Phi}}^{\sigma\sigmap}_\star \pd^2_{\sigma\sigmap} \vec{\phi}
+ \vec{\Gamma} \vec{\phi} \notag
\\ & -i(\tilde{\vec{\mcc{L}}}_{\star\sigma\sigmap} \tilde{\varrho}^{\sigma} \tilde{\varrho}^{\sigmap}
+ \tilde{\vec{\mcc{M}}}_{\star\sigma} \tilde{\varrho}^\sigma
+ \vec{M}_\star - \vec{U}_\star) \vec{\phi}.\label{eq:parab}
\end{align}
Here, we introduced the notation
\begin{gather}
\pd_\zeta \equiv \frac{\pd}{\pd \zeta}, \quad
\pd_\sigma \equiv \frac{\pd}{\pd \varrho^\sigma}, \quad
\pd_{\sigmap} \equiv \frac{\pd}{\pd \varrho^{\sigmap}}
\end{gather}
and the following coefficients:
\begin{gather}
\tilde{\vec{\Phi}}^{\sigma\sigmap}_\star =
\tilde{\mc{X}}_\star^\sigma{}_\alpha\tilde{\mc{X}}_\star ^{\sigmap}{}_\beta H_\star^{|\alpha\beta} \mathbb{1},
\notag\\
\vec{\Gamma} = \vec{\Xi}^+_\star \vec{D}_A(\vec{X} + \vec{\varrho}, \vec{K} + \vec{\dk}) \vec{\Xi}_\star,
\notag\\
\vec{U}_\star = (
H_{\star|\alpha} \vec{\Xi}_\star^+ \vec{\Xi}_\star^{|\alpha}
-H_\star^{|\alpha} \vec{\Xi}_\star^+ \vec{\Xi}_{\star|\alpha}
+ \vec{\Xi}_\star^{+|\alpha} \vec{D}_H \vec{\Xi}_{\star|\alpha}
)_A.\notag
\end{gather}
(As a reminder, the index $A$ in the latter formula denotes the anti-Hermitian part.) Also,
\begin{multline}
\tilde{\vec{\vartheta}}_\star^\sigma{}_{\sigmap} =
[
H^{|\alpha}_{\star|\beta}
- (V_{\star\gamma}/V^2_\star)\,H^{|\alpha\gamma}_\star H_{\star|\beta}
]
\tilde{\mc{X}}^{\sigma}_\star{}_\alpha \mc{X}^\beta_\star{}_{\sigmap}\mathbb{1}
\\- (\tilde{\vec{\mc{X}}}_\star \vec{\mc{Y}}_{\star\sigmap}  \tilde{\vec{\mc{X}}}_\star \vec{V}_\star)^\sigma,
\end{multline}
the matrix $\tilde{\vec{\vartheta}}_\star^\sigma{}_{\sigma}$ is obtained from here by setting $\sigmap = \sigma$ and summing over $\sigma$ accordingly, and
\begin{gather}\label{eq:Y}
\vec{\mc{Y}}_{\star\sigma} = \left(
\begin{array}{ccc}
[(\tilde{\vec{e}}_{\star\sigma})^1]' & 0 & 0 \\[3pt]
[(\tilde{\vec{e}}_{\star\sigma})^2]' & 0 & 0 \\[3pt]
[(\tilde{\vec{e}}_{\star\sigma})^3]' & 0 & 0
\end{array}
\right).
\end{gather}
Finally,
\begin{multline}\label{eq:TaylorL}
\tilde{\vec{\mcc{L}}}_{\star\sigma\sigmap} =
\frac{1}{2}
\bigg(
H_{\star|\alpha\beta}
- \frac{2V_{\star\gamma}}{V^2_\star}
\,H_{\star|\alpha}^{|\gamma}H_{\star|\beta}
\\
+ \frac{V_{\star\gamma}V_{\star\delta}}{V^4_\star}
\,H^{|\gamma\delta}_\star H_{\star|\alpha}H_{\star|\beta}
\bigg)
\mc{X}_\star^\alpha{}_{\sigma} \mc{X}_\star^\beta{}_{\sigmap} \mathbb{1}.
\end{multline}

Note that the structure of the two-mode quasioptical equation \eq{eq:parab} is the same as that of the single-mode equation in Paper~II except for the following: (i) there are additional terms $\tilde{\vec{u}}_\star$, $\smash{\tilde{\vec{\mcc{M}}}_{\star}}$, and $\vec{M}_\star$; and (ii) the coefficients are matrices rather than scalars. In particular, $\vec{U}$ and $\vec{\Gamma}$ are generally non-diagonal and thus can cause mode conversion. Also note that, like in Paper~II, one finds the following corollary:
\begin{gather}
\frac{\dd P}{\dd \zeta} = 2\int (\vec{a}^\dag \vec{\Gamma} \vec{a})\,d^2\tilde{\varrho},
\end{gather}
where $P \doteq \int |\vec{\phi}|^2\,d^2\tilde{\varrho}$ equals, up to a constant coefficient, the energy flux carried by the beam. This shows that the total energy flux of the beam is conserved when $\vec{\Gamma} = 0$.

\subsection{Polarization angles}
\label{sec:alpbet}

\begin{figure}
\begin{center}
\includegraphics[width=8.2cm,clip]{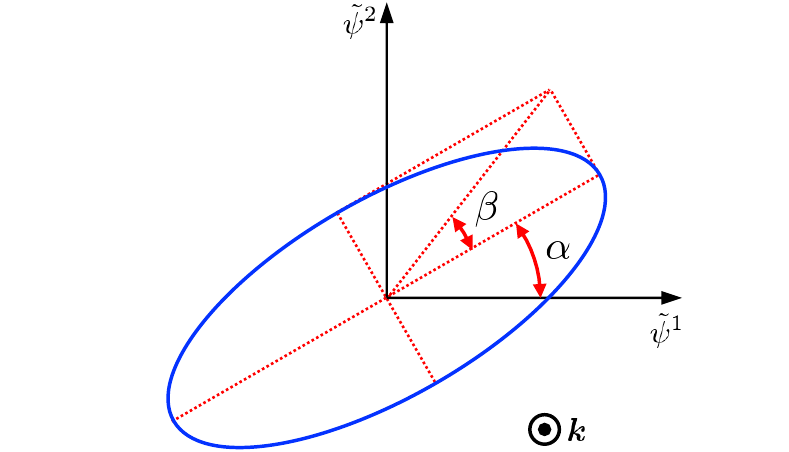}
\caption{Explanation of the polarization angles $\alpha$ and $\beta$. \shout{Specifically,} $\alpha$ denotes the rotation angle of the major axis of the polarization ellipse, and $\beta$ determines the ellipticity.}
\label{fig:alpbet}
\end{center}
\end{figure}

To facilitate benchmarking of \parade, we also introduce the electromagnetic-field parametrization in terms of the polarization angles \cite{book:born65} used in the 1DFW code presented in \Ref{ref:kubo15}. The two components of the field envelope $\vec{\psi}$ projected on the transverse space $\{\tilde{\varrho}^1, \tilde{\varrho}^2\}$ can be expressed as follows:
\begin{gather}\label{eq:ab2psi}
\left(
\begin{array}{c}
\tilde{\psi}^1_\star
\\
\tilde{\psi}^2_\star
\end{array}
\right)
= |\vec{\psi}_\star|
\left(
\begin{array}{c}
\cos \alpha \cos \beta - i \sin \alpha \sin \beta
\\
\sin \alpha \cos \beta + i \cos \alpha \sin \beta
\end{array}
\right).
\end{gather}
The geometrical meaning of the polarization angles $\alpha$ and $\beta$ can be understood from \Fig{fig:alpbet}. Explicitly, these angles can be expressed through $\vec{\psi}$ as \cite{book:born65}
\begin{gather}\label{eq:psi2ab}
\alpha = \frac{1}{2} \tan^{-1}
\bigg(
\frac{2|\tilde{\psi}^1_\star| |\tilde{\psi}^2_\star|}{|\tilde{\psi}^1_\star|^2 - |\tilde{\psi}^2_\star|^2} \cos \delta
\bigg),
\\
\beta  = \frac{1}{2} \sin^{-1}
\bigg(
\frac{2|\tilde{\psi}^1_\star| |\tilde{\psi}^2_\star|}{|\tilde{\psi}^1_\star|^2 + |\tilde{\psi}^2_\star|^2} \sin \delta
\bigg)
\end{gather}
(note the difference in the signs in the denominators), where $\delta \doteq \arg (\tilde{\psi}^2_\star) - \arg (\tilde{\psi}^1_\star)$.
In the following \parade simulations, the initial $\vec{a}$ is prescribed by prescribing $\alpha$ and $\beta$. Specifically, $\alpha$ and $\beta$ determine $\tilde{\psi}^\mu$ via \Eq{eq:ab2psi}, and \Eq{eq:apsi} yields $a^s = \eta^{s*}{}_\alpha \mc{X}^\alpha{}_\mu \tilde{\psi}^\mu$.

\section{Simulation results}
\label{sec:sim}

Here, we present \parade simulations of mode-converting wave beams with the focus on two effects: (i)~the mode-amplitude transformation during the O--X conversion caused by the magnetic-field shear and~(ii) splitting of beams that consist of multiple modes. The simulation algorithm is the same as the one used in \parade for single-mode waves in Paper~II. \shout{Also like in Paper~II, all simulations were done on a laptop with Intel Core\textsuperscript{TM}~i7-7660U processor and took only a few seconds to run, as further specified in the figure captions.} For $\vec{D}$, we assume the dispersion tensor of collisionless cold electron plasma for simplicity, so there is no dissipation.

\subsection{Mode conversion in a sheared magnetic field}
\label{sec:mc}

\subsubsection{Comparison with uncoupled-mode simulations}
\label{sec:U}

As mentioned in \Sec{sec:qo}, mode conversion in the vector equation \eq{eq:parab} is governed by non-diagonal matrices $\vec{U}_\star$ and $\vec{\Gamma}$. In the cold-plasma model, $\vec{\Gamma}$ is zero but $\vec{U}_\star$ is generally non-negligible. In order to illustrate the effect of $\vec{U}_\star$ on the polarization state of wave beams in low-density plasmas, we compared \parade simulations using \Eq{eq:parab} with those using the single-mode scheme. This scheme is described in Paper~II, and it is also similar to other existing quasioptical models \cite{ref:mazzucato89, ref:nowak93, ref:peeters96, ref:farina07, ref:pereverzev98, ref:poli01b, ref:poli01, ref:poli18, ref:balakin08b, ref:balakin07a, ref:balakin07b} in that it ignores the coupling between the electromagnetic modes.

As an example, we chose the initial wave to be a pure O mode. Also, the assumed geometry is as follows. We introduce the standard notation $\{x, y, z\}$ for the laboratory coordinates. The origin is chosen to be the RR starting point, and $\vec{e}_z$, which is the unit vector along the $z$ axis, is chosen to be the orientation of the RR initial wave vector. Then, we assume a slab geometry with electron density
\begin{gather}\label{eq:slabn}
n_e(z) = n_0 \exp\left( \frac{z - z_0}{L_n} \right)
\end{gather}
and magnetic field
\begin{gather}
\label{eq:slabB}
\left(
\begin{array}{c}
B_x(z)\\
B_y(z)\\
B_z
\end{array}
\right)
= B_0
\left(
\begin{array}{c}
\sin \theta_o\,\cos \left( \theta_s + 2 \pi z / L_b \right)\\
\sin \theta_o\,\sin \left( \theta_s + 2 \pi z / L_b \right)\\
\cos \theta_o
\end{array}
\right).
\end{gather}
The parameters are $n_0 = 1.0 \times 10^{18}$~m$^{-3}$, $z_0 = 0.9$~m, $L_n = 0.9$~m, $B_0 = 0.4$~T, $\theta_o = 80.0^\circ$, $\theta_s = 80.0^\circ$, and $L_b = 0.9$~m. The polarization angles are chosen to be $\alpha = 80.0^\circ$ and $\beta = -27.0^\circ$. We also adopt the initial beam profile as Gaussian \cite{foot:paper2, book:yariv} with the focal lengths $\mc{Z}_1 = \mc{Z}_2 = 4.0$~m and waist sizes $w_{0,1} = w_{0,2} = 5.0$~cm. The wave frequency is $f = 77.0$~GHz, which corresponds to the vacuum wavelength $\lambda_0 \approx 4$~mm.

\begin{figure}
\begin{center}
\includegraphics[width=8.2cm,clip]{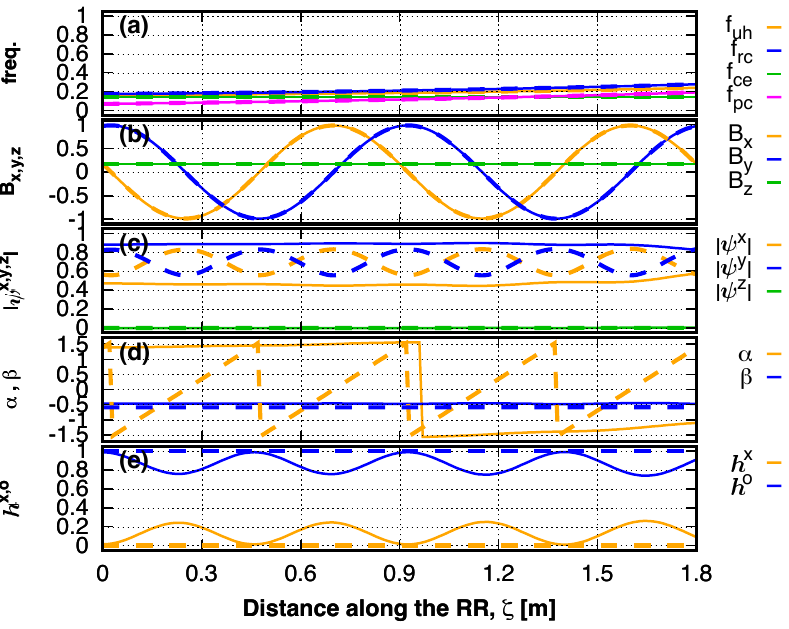}
\end{center}
\caption{Simulations of the vector-beam propagation parallel to the density gradient in cold electron plasma with a sheared magnetic field [\Eqs{eq:slabn} and \eq{eq:slabB}]. The parameters are as specified in \Sec{sec:U}. The solid lines show the \parade predictions based on \Eq{eq:parab}, which resolves mode conversion. The dashed lines show the \parade predictions based on the single-mode scheme (Paper~II). (a) shows the key frequencies on the RR trajectory, namely, the upper-hybrid frequency $f_{\rm uh}$, the right-cutoff frequency $f_{\rm rc}$, the electron cyclotron frequency  $f_{\rm ce}$, and the plasma frequency $f_{\rm pc}$, all in units $f$. (b) shows the components of the magnetic field $(B_x, B_y, B_z)$ in units of its local strength $|\vec{B}|$. (c) shows the absolute values of the individual components of $\vec{\psi}$ \shout{on the RR}. (d) shows the polarization angles $\alpha$ and $\beta$. (e) shows the relative intensities of the O and X waves [\Eq{eq:h}]. \shout{The computing time is approximately $4.5$~s.}}
\label{fig:U}
\end{figure}

The simulation results are presented in \Fig{fig:U}. It is seen that, when the vector model \eq{eq:parab} is used (solid lines), the variation of the polarization angles is slow and overall small [\Fig{fig:U}(d)], because the plasma density is low ($f_{\rm pc} < 0.2$ in units $f$). Correspondingly, the relative intensities of the O and X waves, which are defined as
\begin{gather}\label{eq:h}
h^s \doteq \frac{\int |a^s|^2\,d^2\tilde{\varrho}}{\int |\vec{a}|^2\,d^2\tilde{\varrho}},
\end{gather}
vary rapidly [\Fig{fig:U}(e)], because of the strong shear of the magnetic field with scale length $L_b = 0.9$~m. This variation illustrates the shear-driven mode conversion. In contrast, in the single-mode scheme (dashed lines), where the mode conversion is not taken into account, the initially-excited O mode remains pure. Then, $h^s$ are fixed and, accordingly, the polarization angles vary rapidly, although the ambient plasma has low density.

\subsubsection{Comparison with the 1DFW code}
\label{sec:FW}

\begin{figure}
\begin{center}
\includegraphics[width=8.2cm,clip]{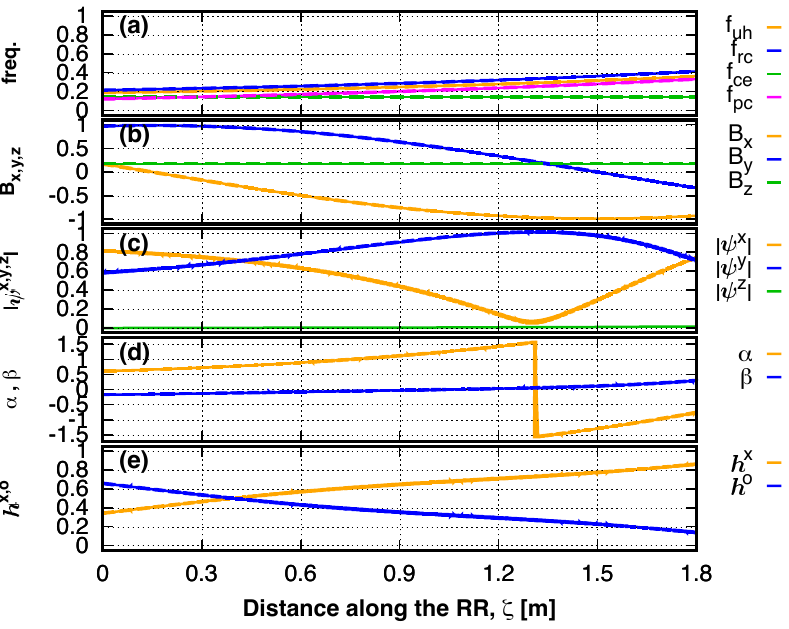}
\end{center}
\caption{Comparison of the simulation results produced by \parade (solid lines) and the 1DFW code (dashed lines). The parameters are the same as in \Fig{fig:U} except $\alpha = 35.0^\circ$, $\beta = -10.0^\circ$, $n_0 = 3.0 \times 10^{18}$~m$^{-3}$, and $L_b = 5.4$~m. \shout{The computing time is approximately $4.5$~s.}}
\label{fig:FW}
\end{figure}

As the second example, we compared predictions of \parade with predictions of the 1DFW code presented in \Ref{ref:kubo15}, so as to verify that the shear-driven mode conversion and smooth variation of the polarization state are modeled accurately. Here, the initial polarization angles are chosen to be $\alpha = 35.0^\circ$ and $\beta = -10.0^\circ$. The simulations are performed in a slab geometry [\Eqs{eq:slabn} and \eq{eq:slabB}] with $n_0 = 3.0 \times 10^{18}$~m$^{-3}$ and $L_b = 5.4$~m; the other parameters are the same as in \Sec{sec:U}. The comparison shows good agreement of the two codes (\Fig{fig:FW}).

\subsubsection{Parameter scan}

\begin{figure}
\begin{center}
\includegraphics[width=8.2cm,clip]{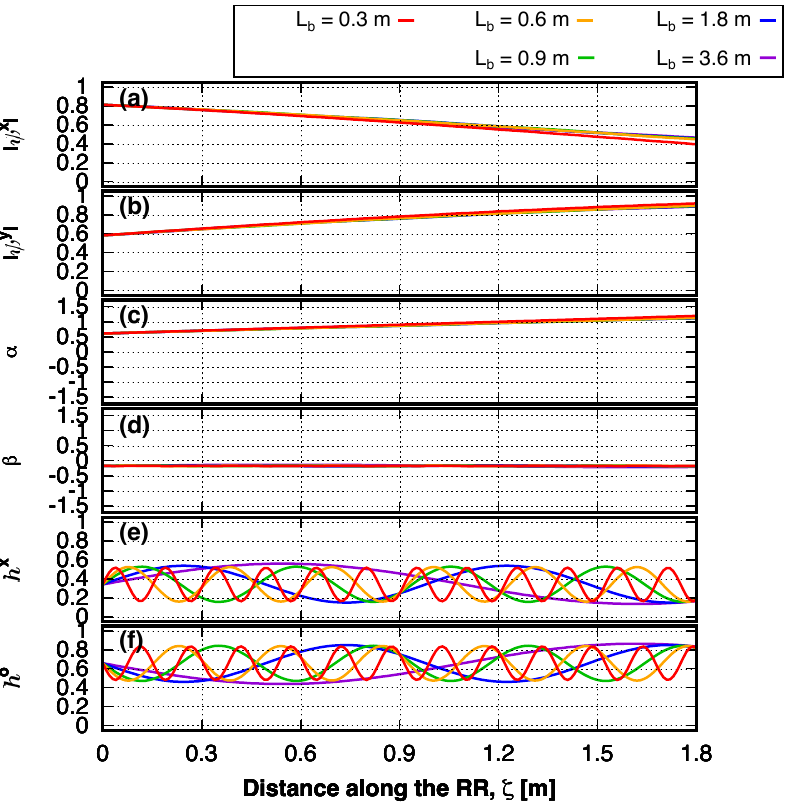}
\end{center}
\caption{\parade simulation results of the O--X mode conversion in a slab geometry with constant density $n = n_0$ and magnetic field given by \Eq{eq:slabB}. The density is fixed, $n_0 = 1.0 \times 10^{18}$~m$^{-3}$, and $L_b$ is scanned in the range between $0.3$~m and $3.6$~m. The other parameters are the same as in \Fig{fig:FW}. (a) and (b) show the absolute values of the two components of $\vec{\psi}$. (c) and (d) show the polarization angles $\alpha$ and $\beta$. (e) and (f) show the relative intensities of the O and X waves [\Eq{eq:h}].}
\label{fig:shear}
\end{figure}

As another example, we consider how the mode conversion is influenced by the magnetic-field shear and by the plasma density. For simplicity, we adopt that the magnetic field is given by \Eq{eq:slabB}, as earlier, whereas the density is constant, $n = n_0$.
First, the O--X mode conversion is simulated with fixed $n_0 = 1.0 \times 10^{18}$~m$^{-3}$ and different scales of the shear, ranging from $L_b = 0.3$~m to $L_b = 3.6$~m (\Fig{fig:shear}). The parameters other than $n_0$ and $L_b$ are kept the same as in \Sec{sec:FW}. Since the low density is assumed ($f_{\rm pc} < 0.2$ in units $f$), the polarization state does not change notably with $L_b$ [\Figs{fig:shear}(c) and (d)]. However, the relative intensities of the O and X waves [\Eq{eq:h}] \textit{do}, because $a^s$ are defined as the projections of $\vec{\psi}$ on the polarization vectors [\Eq{eq:apsi}] that are linked to $\vec{B}$. As seen in \Figs{fig:shear}(e) and (f), smaller $L_b$ corresponds to shorter periods of the energy exchange between the O and X waves. Specifically, the mode conversion occurs twice per rotation of the magnetic field in the $(x, y)$ plane.

\begin{figure}
\begin{center}
\includegraphics[width=8.2cm,clip]{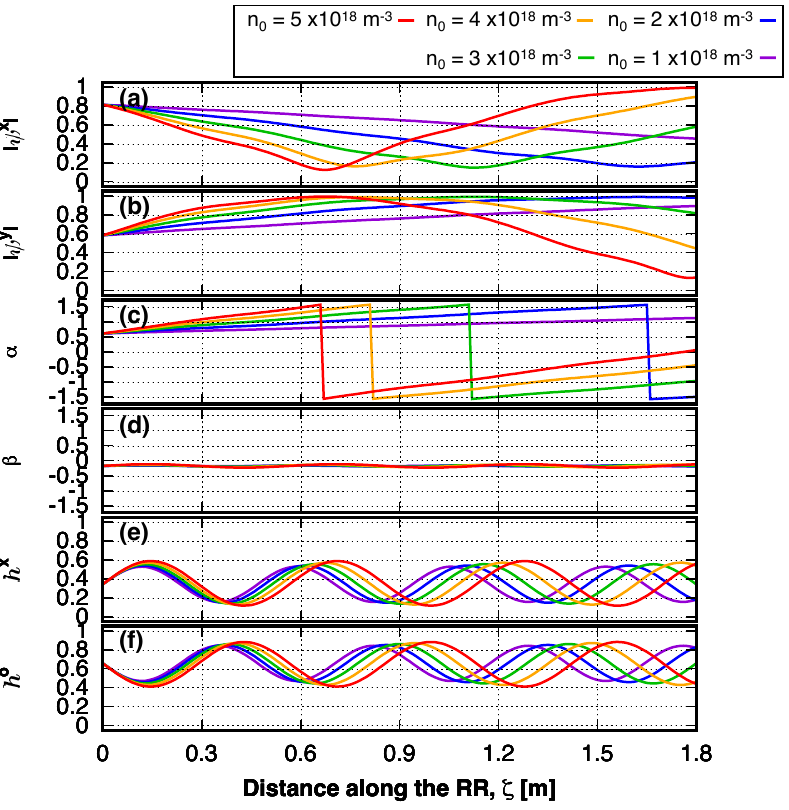}
\end{center}
\caption{Same as in \Fig{fig:shear} except $L_b = 0.9$~m is fixed and $n_0$ is scanned in the range between $1.0 \times 10^{18}$~m$^{-3}$ to $5.0 \times 10^{18}$~m$^{-3}$ with the increment $1.0 \times 10^{18}$~m$^{-3}$.}
\label{fig:dense}
\end{figure}

We also simulated the O--X mode conversion for $n_0$ scanned in the range between $1.0 \times 10^{18}$~m$^{-3}$ to $5.0 \times 10^{18}$~m$^{-3}$ at fixed $L_b = 0.9$~m. The other parameters were kept the same as in \Sec{sec:FW}. As seen in \Fig{fig:dense}, larger $n_0$ corresponds to stronger variations of the polarization angles [\Figs{fig:dense}(c) and (d)] and slower variations of the relative intensities [\Figs{fig:dense}(e) and (f)]. This tendency predicted by \parade is anticipated, because the two electromagnetic eigenmodes are nonresonant at sufficiently dense plasma but become resonant at low densities, where they both have the refraction indexes close to unity.
The influence of the shear [\Figs{fig:shear}(e) and (f)] and plasma density [\Figs{fig:dense}(e) and (f)] on the relative intensities is particularly relevant in practice due to the fact that magnetically-confined fusion plasmas have inhomogeneous shear and nonzero inhomogeneous density even outside the edge. Specific applications of \parade and comparison with 1DFW simulations (where refraction is neglected) are reported in \Ref{foot:itc27}.

\subsection{Weak splitting of mode-converting beams}
\label{sec:split}

\begin{figure*}
\begin{center}
\includegraphics[width=15.0cm,clip]{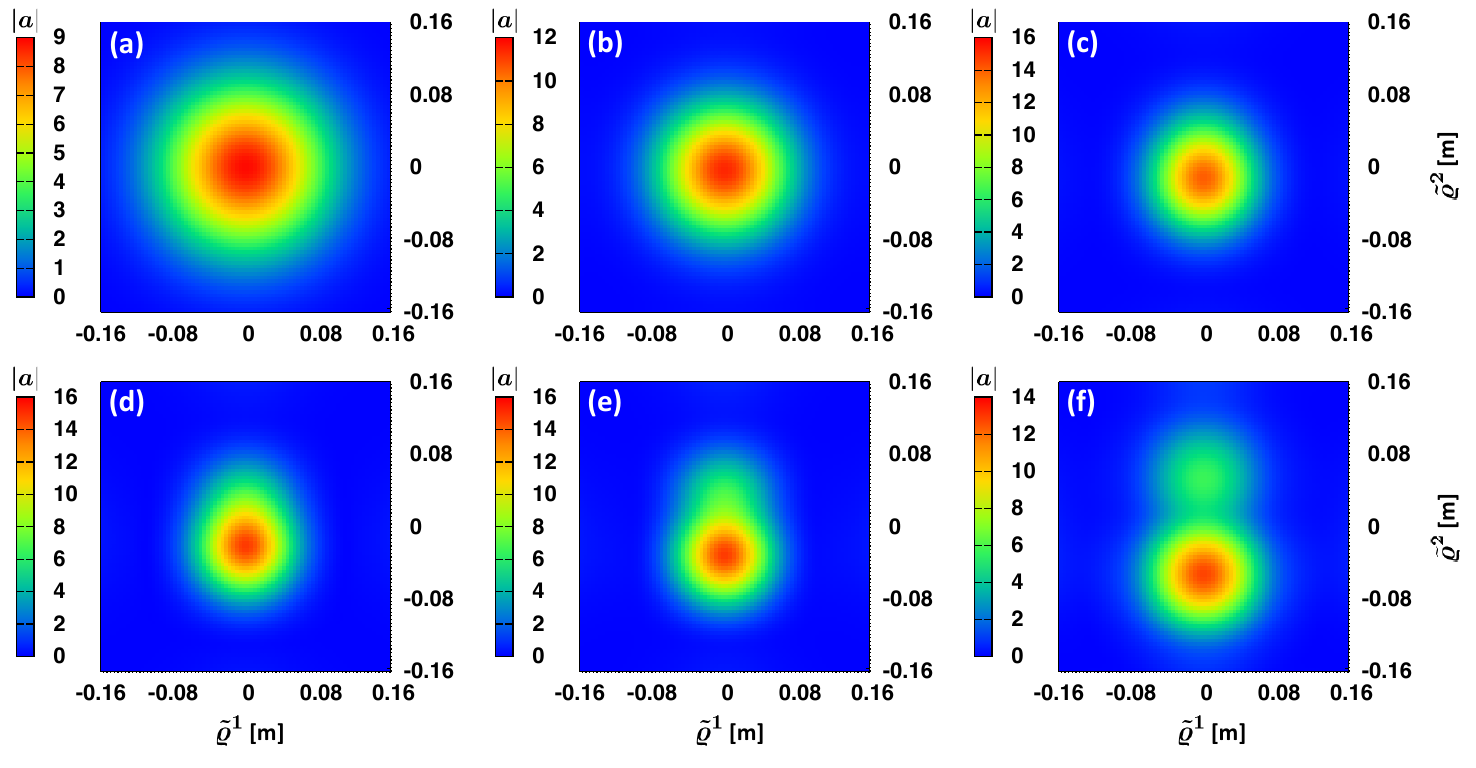}
\caption{\parade simulation of the quasioptical-beam splitting in a slab geometry [\Eqs{eq:nB21} and \eq{eq:nB22}]. The parameters are as specified in \Sec{sec:split}. The transverse profile of the field amplitude $|\vec{a}| \approx |\vec{\psi}|$ at various locations along the beam trajectory: (a) $\zeta = 0.0$~m, (b) $\zeta = 1.0$~m, (c) $\zeta = 2.0$~m, (d) $\zeta = 2.5$~m, (e) $\zeta = 3.0$~m, and (f) $\zeta = 4.0$~m. \shout{The computing time is approximately $9$~s.} The figures illustrate the gradual splitting of the original beam into O-mode and X-mode beams propagating along separate ray trajectories. Here, the splitting is weak, so both beams remain close enough to their common RR, which makes it possible to simulate the process with a quasioptical model. At later times, each beam can be simulated with \parade independently as a single-mode beam.}
\label{fig:prof}
\end{center}
\end{figure*}

\begin{figure*}
\begin{center}
\includegraphics[width=15.0cm,clip]{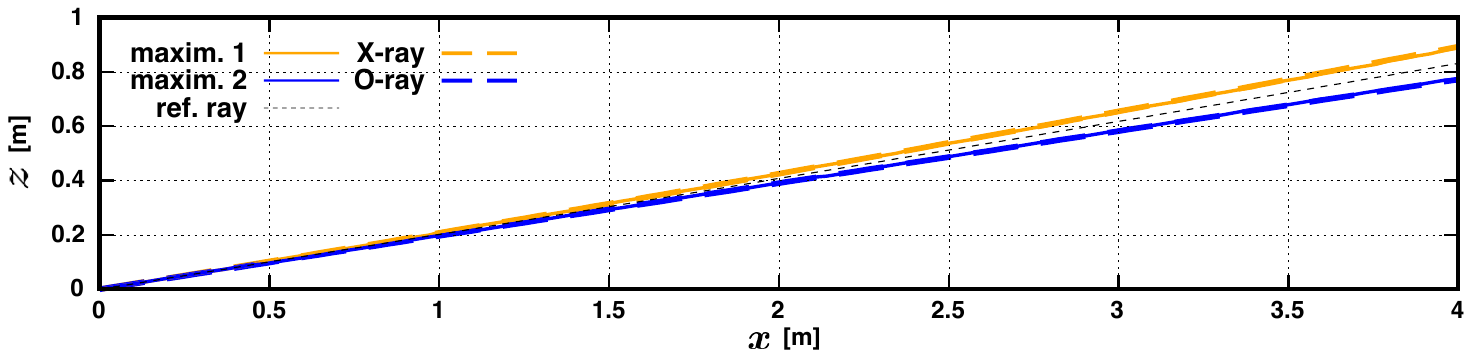}
\caption{The locations of the amplitude maxima in a two-mode beam under the same conditions as in \Fig{fig:prof}: a \parade simulation (solid lines) versus a ray-tracing simulation (long-dashed lines); also shown is the RR trajectory (short-dashed line).}
\label{fig:split}
\end{center}
\end{figure*}

\parade is also advantageous in that it can efficiently model splitting of multi-mode beams. To demonstrate this capability, we performed numerical simulations in a slab geometry with
\begin{gather}
n = n_0 \exp \bigg[ - \frac{(x - x_0)^2}{L_n^2} \bigg],\label{eq:nB21}
\\
\vec{B} = \vec{e}_z B_0 \exp \bigg[ - \frac{(x - x_0)^2}{L_b^2} \bigg].\label{eq:nB22}
\end{gather}
Here, $n_0 = 1.0 \times 10^{19}$~m$^{-3}$, $x_0 = 4.0$~m, $L_n = 4.0$~m, $B_0 = 1.0$~T, $L_b = 4.0$~m, and $\vec{e}_z$ is the unit vector along the $z$ axis. (The orientation of \shout{the} magnetic field $\vec{B} / |\vec{B}|$ is chosen to be constant here to suppress the shear-driven mode conversion.) The origin is chosen to be the RR starting point, and $\{x, y, z\} = \{1.0, 0.0, 0.2\}$~m is chosen to be the target point, which is used to fix the orientation of the RR initial wave vector. The initial polarization angles are $\alpha = 10.0^\circ$ and $\beta = -30.0^\circ$. Also, for the initial beam profile, we adopt a Gaussian profile \cite{foot:paper2, book:yariv} with the focal lengths $\mc{Z}_1 = \mc{Z}_2 = 3.0$~m and waist sizes $w_{0,1} = w_{0,2} = 5.0$~cm. The wave frequency is chosen to be $f = 77.0$~GHz.
Figure~\ref{fig:prof} shows the evolution of the transverse profile of $|\vec{a}|$ at different locations along the RR. One can see the gradual splitting of the original beam into O-mode and X-mode beams propagating along separate ray trajectories. (Note that a single RR is used for this simulation, in contrast to single-mode simulations, where each mode would have its own RR.)
Figure~\ref{fig:split} shows the trajectories of the locations of the amplitude maxima. For comparison, this figure also shows the trajectories obtained from ray-tracing simulations, where O and X waves are modeled as independent. It is seen that \parade's quasioptical simulations are in good agreement with conventional ray tracing.
Importantly, such quasioptical modeling of a two-mode beam is adequate only as long as the group velocities of the O and X waves remain close enough to each other; otherwise the ordering \eq{eq:eps} cannot be maintained. (That is why we consider only weak splitting here.) However, by the time the two group velocities become very different, the O and X waves also become nonresonant and thus independent. Such waves can also be modeled with \parade, except the single-mode algorithm \cite{foot:paper2} must be used instead.

\section{Conclusions}
\label{sec:conc}

This work continues a series of papers where we propose a new code \parade for quasioptical modeling of electromagnetic beams with and without mode conversion. The general theoretical model underlying \parade and its application to single-mode beams were presented earlier \cite{foot:paper1, foot:paper2}. Here, we apply \parade to produce the first quasioptical simulations of mode-converting beams. We also demonstrate that \parade can model splitting of two-mode beams. The numerical results produced by \parade show good agreement with those of one-dimensional full-wave simulations and also with conventional ray tracing \shout{to the extent those are applicable}.

\section{Acknowledgments}
The work was supported by the U.S. DOE through Contract No. DE-AC02-09CH11466. The work was also supported by JSPS KAKENHI Grant Number JP17H03514.


\begin{thebibliography}{10}

\bibitem{ref:tsujimura15}
T.~I. Tsujimura, S.~Kubo, H.~Takahashi, R.~Makino, R.~Seki, Y.~Yoshimura,
  H.~Igami, T.~Shimozuma, K.~Ida, C.~Suzuki, M.~Emoto, M.~Yokoyama,
  T.~Kobayashi, C.~Moon, K.~Nagaoka, M.~Osakabe, S.~Kobayashi, S.~Ito,
  Y.~Mizuno, K.~Okada, A.~Ejiri, T.~Mutoh, and {the LHD Experiment Group}, {\it
  Development and application of a ray-tracing code integrating with 3D
  equilibrium mapping in LHD ECH experiments\/}, Nucl. Fusion {\bf 55}, 123019
  (2015).

\bibitem{ref:marushchenko14}
N.~B. Marushchenko, Y.~Turkin, and H.~Maassberg, {\it Ray-tracing code {TRAVIS}
  for {ECR} heating, {EC} current drive and {ECE} diagnostic\/}, Comput. Phys.
  Commun. {\bf 185}, 165 (2014).

\bibitem{ref:mazzucato89}
E.~Mazzucato, {\it Propagation of a Gaussian beam in a nonhomogeneous
  plasma\/}, Phys. Fluids B {\bf 1}, 1855 (1989).

\bibitem{ref:nowak93}
S.~Nowak and A.~Orefice, {\it Quasioptical treatment of electromagnetic
  Gaussian beams in inhomogeneous and anisotropic plasmas\/}, Phys. Fluids B
  {\bf 5}, 1945 (1993).

\bibitem{ref:peeters96}
A.~G. Peeters, {\it Extension of the ray equations of geometric optics to
  include diffraction effects\/}, Phys. Plasmas {\bf 3}, 4386 (1996).

\bibitem{ref:farina07}
D.~Farina, {\it A quasi-optical beam-tracing code for electron cyclotron
  absorption and current drive: GRAY\/}, Fusion Sci. Tech. {\bf 52}, 154
  (2007).

\bibitem{ref:pereverzev98}
G.~V. Pereverzev, {\it Beam tracing in inhomogeneous anisotropic plasmas\/},
  Phys. Plasmas {\bf 5}, 3529 (1998).

\bibitem{ref:poli01b}
E.~Poli, G.~V. Pereverzev, A.~G. Peeters, and M.~Bornatici, {\it {EC} beam
  tracing in fusion plasmas\/}, Fusion Eng. Des. {\bf 53}, 9 (2001).

\bibitem{ref:poli01}
E.~Poli, A.~G. Peeters, and G.~V. Pereverzev, {\it {TORBEAM}, a beam tracing
  code for electron-cyclotron waves in tokamak plasmas\/}, Comput. Phys.
  Commun. {\bf 136}, 90 (2001).

\bibitem{ref:poli18}
E.~Poli, A.~Bock, M.~Lochbrunner, O.~Maj, M.~Reich, A.~Snicker, A.~Stegmeir,
  F.~Volpe, N.~Bertelli, R.~Bilato, G.~D. Conway, D.~Farina, F.~Felici,
  L.~Figini, R.~Fischer, C.~Galperti, T.~Happel, Y.~R. Lin-Liu, N.~B.
  Marushchenko, U.~Mszanowski, F.~M. Poli, J.~Stober, E.~Westerhof, R.~Zille,
  A.~G. Peeters, and G.~V. Pereverzev, {\it {TORBEAM} 2.0, a paraxial beam
  tracing code for electron-cyclotron beams in fusion plasmas for extended
  physics applications\/}, Comput. Phys. Commun. {\bf 225}, 36 (2018).

\bibitem{ref:balakin08b}
A.~A. Balakin, M.~A. Balakina, and E.~Westerhof, {\it ECRH power deposition
  from a quasi-optical point of view\/}, Nucl. Fusion {\bf 48}, 065003 (2008).

\bibitem{ref:balakin07a}
A.~A. Balakin, M.~A. Balakina, G.~V. Permitin, and A.~I. Smirnov, {\it
  Quasi-optical description of wave beams in smoothly inhomogeneous anisotropic
  media\/}, J. Phys. D: Appl. Phys. {\bf 40}, 4285 (2007).

\bibitem{ref:balakin07b}
A.~A. Balakin, M.~A. Balakina, G.~V. Permitin, and A.~I. Smirnov, {\it Scalar
  equation for wave beams in a magnetized plasma\/}, Plasma Phys. Rep. {\bf
  33}, 302 (2007).

\bibitem{ref:kubo15}
S.~Kubo, H.~Igami, T.~I. Tsujimura, T.~Shimozuma, H.~Takahashi, Y.~Yoshimura,
  M.~Nishiura, R.~Makino, and T.~Mutoh, {\it Plasma interface of the EC waves
  to the LHD peripheral region\/}, AIP Conf. Proc. {\bf 1689}, 090006 (2015).

\bibitem{my:xo}
I.~Y. Dodin, D.~E. Ruiz, and S.~Kubo, {\it Mode conversion in cold low-density
  plasma with a sheared magnetic field\/}, Phys. Plasmas {\bf 24}, 122116
  (2017).

\bibitem{phd:ruiz17}
D.~E. Ruiz, {\it Geometric theory of waves and its applications to plasma
  physics\/}, Ph.D. Thesis, Princeton Univ. (2017), arXiv:1708.05423.

\bibitem{my:covar}
D.~E. Ruiz and I.~Y. Dodin, {\it Extending geometrical optics: A Lagrangian
  theory for vector waves\/}, Phys. Plasmas {\bf 24}, 055704 (2017).

\bibitem{my:qdirac}
D.~E. Ruiz and I.~Y. Dodin, {\it Lagrangian geometrical optics of nonadiabatic
  vector waves and spin particles\/}, Phys. Lett. A {\bf 379}, 2337 (2015).

\bibitem{my:qdiel}
D.~E. Ruiz and I.~Y. Dodin, {\it First-principles variational formulation of
  polarization effects in geometrical optics\/}, Phys. Rev. A {\bf 92}, 043805
  (2015).

\bibitem{foot:paper1}
I.~Y. Dodin, D.~E. Ruiz, K. Yanagihara, Y. Zhou, and S. Kubo, {\it Quasioptical
  modeling of wave beams with and without mode conversion: I.~Basic theory},
  arXiv:1901.00268.

\bibitem{foot:paper2}
K. Yanagihara, I.~Y. Dodin, and S. Kubo, {\it Quasioptical modeling of wave
  beams with and without mode conversion: II.~Numerical simulations of
  single-mode beams}, submitted in parallel.

\bibitem{ref:bliokh15}
K.~Y. Bliokh, F.~J. Rodr{\'{\i}}guez-Fortu{\~{n}}o, F.~Nori, and A.~V. Zayats,
  {\it Spin-orbit interactions of light\/}, Nat. Photonics {\bf 9}, 796 (2015).

\bibitem{foot:marius}
M.~A. Oancea, C.~F. Paganini, J. Joudioux, and L. Andersson, \textit{An
  overview of the gravitational spin Hall effect}, arXiv:1904.09963.

\bibitem{foot:itc27}
K. Yanagihara, S. Kubo, T.~I. Tsujimura, and I.~Y. Dodin, {\it Mode purity of
  electron cyclotron waves after their passage through the peripheral plasma in
  the Large Helical Device}, \shout{to appear in Plasma Fusion Res.}.

\bibitem{book:stix}
T.~H. Stix, {\it Waves in Plasmas\/} (AIP, New York, 1992).

\bibitem{book:born65}
M.~Born and E.~Wolf, {\it Principles of Optics\/} (Pergamon Press, Oxford,
  1965).

\bibitem{book:yariv}
A.~Yariv, {\it Quantum Electronics\/} (Wiley, New Jersey, 1967).

\end{thebibliography}

\end{document}